\documentstyle[prd,aps,axodraw,
preprint,
]{revtex}

\begin{document}

\newcommand{\TeV}{\,{\rm TeV}}
\newcommand{\GeV}{\,{\rm GeV}}
\newcommand{\MeV}{\,{\rm MeV}}
\newcommand{\keV}{\,{\rm keV}}
\newcommand{\eV}{\,{\rm eV}}
\def\ap{\approx}
\newcommand{\bea}{\begin{eqnarray}}
\newcommand{\eea}{\end{eqnarray}}
\def\beq{\begin{equation}}
\def\eeq{\end{equation}}
\def\haf{\frac{1}{2}}
\def\lpp{\lambda''}
\def\ccg{\cal G}
\def\slash#1{#1\!\!\!\!\!\!/}
\def\rpv{{\slash{R_p}\;}}
\def\u{{\cal U}}

\setcounter{page}{1}
\draft
\preprint{SCIPP-00/16, KAIST-TH 00/08, LANCS-TH/0006}

\title{Lepton Flavor Violation and Bilinear R-parity Violation}
 
\author{Kiwoon Choi$^{a,b}$, Eung Jin Chun$^{c,d}$ and Kyuwan Hwang$^b$}

\address{$^a$Santa Cruz Institute for Particle Physics, Santa Cruz,
CA 95064, U. S. A. \\
$^b$Korea Advanced Institute of Science and Technology, Taejeon 305-701,
Korea \\
$^c$Korea Institute for Advanced Study, Seoul 130-012, Korea\\
$^d$Department of Physics, Lancaster University, Lancaster LA1 4YB, UK
}


\maketitle

\begin{abstract}
We examine some flavor changing  processes such as
rare leptonic decays of the long-lived neutral kaon, muon-electron
conversion in nuclei and radiative muon decay which are
induced by the combined effects of  bilinear and trilinear 
R-parity violations.
These processes are used  to put strong constraints on 
certain products of  the bilinear and trilinear  couplings.
We also discuss the constraints on R-parity violation from
neutrino masses and compare them with the constraints
from flavor changing decay processes.
Large range of parameter space
satisfying the constraints from neutrino masses can be excluded by
the flavor changing processes considered in this paper, and also
vice versa.

\end{abstract}



R-parity violation in the minimal supersymmetric standard model
can be an interesting  source of lepton and quark flavor violation.
Typically one obtains bounds on R-parity violating couplings
from the non-observation of such flavor violating phenomena in experiments.
There are a vast of literatures addressing this issue \cite{review}.
However most of them have focused on the effects of trilinear R-parity
violation alone.
In  generic context, both bilinear and trilinear R-parity violations
exist and they are independent from each other.
Our focus in this paper is on the effects of bilinear R-parity violation
on some flavor changing  processes such as $K_L\to e_i \bar{e}_j$,
$K^+ \to \pi^+ \nu_i \bar{\nu}_j$, $\mu$--$e$ conversion in nuclei, 
and $\mu \to e\gamma$.

These  processes  can arise from tree or one-loop diagrams
involving two insertions of the trilinear R-parity violating couplings
$\lambda$ and $\lambda^{\prime}$.
The resulting bounds on $\lambda\lambda$ and $\lambda^{\prime}
\lambda^{\prime}$ are obtained 
in the literature for 
$K_L\to e_i \bar{e}_j$ \cite{CR}, 
$K^+ \to \pi^+ \nu_i \bar{\nu}_j$ \cite{AG},
$\mu$--$e$ conversion in nuclei \cite{huitu,Faessler}, 
and $\mu\to e\gamma$ \cite{CH}.
Flavor changing decays of the neutral kaon
and also the $\mu$--$e$ conversion in nuclei
can be achieved by tree diagrams without any insertion of
small fermion mass, thereby put very strong bounds of order
$10^{-6}\sim 10^{-8}$ on $\lambda\lambda$ and 
$\lambda^{\prime}\lambda^{\prime}$. 
However one-loop diagrams for $\mu \to e\gamma$ 
need  additional chirality flip by small fermion mass.
The decay amplitude is then suppressed by  $m_\mu/m_{\tilde{f}}$ 
with $m_{\tilde{f}}$ being  a typical sfermion mass. This  makes 
the  bounds on $\lambda\lambda$ and 
$\lambda^{\prime}\lambda^{\prime}$ from $\mu\rightarrow
e\gamma$  looser, which are 
of order $10^{-4}$.

In the presence of bilinear R-parity violation,
there are additional contributions involving both 
bilinear and trilinear couplings.  
The importance  of such contribution to fermion electric
dipole moments have been discussed recently in Refs.\cite{CCH,KK}.
In this case, the chirality structure of diagrams  differs from
the case with trilinear couplings only. For instance,
there are  one-loop diagrams
for $\mu\rightarrow e\gamma$ without having the chirality
flip by  small fermion mass. As summarized in Table II, this leads  
to strong bounds of order $10^{-7}$ on $\epsilon\lambda$
where $\epsilon$ is a dimensionless parameter measuring the size
of bilinear R-parity violation.
(See Table II.)
On the other hand, tree diagrams for neutral kaon decays 
have to come with another chirality flip by  small fermion mass,
thereby lead to looser bounds of order $10^{-4}\sim 10^{-5}$ 
on $\epsilon\lambda$ and  $\epsilon\lambda^{\prime}$
as summarized in Table I.

Since R-parity violation is severely constrained 
by neutrino masses \cite{hs}, it is quite tempting
to compare the bounds of Tables I and II with those
from neutrino masses \cite{haug,abada,Lee}.
The recent Super-Kamiokande \cite{skam} and other neutrino
oscillation data can be well explained by assuming small Majorana
neutrino masses: $m_{\nu}\sim 10^{-1}$ eV or less.
It has been argued that the existing neutrino data
suggest that the maximal values of {\it all} $3\times 3$ neutrino mass
matrix elements do not exceed 1 eV \cite{haug}.
Here we take this  as a real bound on neutrino
masses and derive the resulting constraints on R-parity violations
which are summarized in Table III.
As we will see, the constraints from neutrino masses and those
of Tables I and II are complementary to each other.
Large range of parameter space satisfying the conditions
for small neutrino masses can be excluded
by the bounds of Tables I and II, and also vice versa.

To proceed, let us first specify the parameter basis for
our analysis. When R-parity conservation is not assumed, the MSSM allows for
renormalizable  lepton number (L) or baryon number (B) violating 
interactions.
In this paper, we ignore B-violating couplings
by simply assuming that they are small enough to ensure
the proton stability.
We will then work in the basis in which 
the standard Yukawa couplings of 
the quarks and leptons  are diagonal  and also
the vacuum expectation  values of sneutrinos vanish \cite{yuval}.
Since this choice is not invariant under the
renormalization group evolution, it must be understood as 
a basis choice for the parameters renormalized at the weak 
scale\footnote{The scale-dependence of choosing basis would  motivate
for a basis-independent prescription for R-parity violating
physics as attempted in \cite{banks}. 
Making a basis-independent analysis
for the processes under consideration is much more involved, so
here we discuss everything in the specific basis with
$\langle\tilde{\nu}_i\rangle=0$ applied for the parameters
renormalized at the weak scale.}. 
In this basis, still the most generic  R-parity 
violating superpotential  is  given by 
\beq
\label{WZ}
\Delta W 
= \epsilon_i {\mu} L_i H_2 + {\lambda}_{ijk}L_iL_jE_k^c
+ {\lambda}'_{ijk} L_iQ_jD^c_k \, ,
\eeq
where  $H_{2}$ is the
Higgs doublet superfield with the hypercharge $Y=1/2$, $L_i$ and 
$Q_i$ are the 
lepton and quark doublet superfields, $E^c_i$ and
$U^c_i,D^c_i$ are the anti-lepton and anti-quark singlet superfields, 
respectively.
Here the dimensionless parameters
$\epsilon_i$ are introduced to parameterize the bilinear R-parity violation
in unit of the supersymmetric  Higgs  mass parameter $\mu$.
Bilinear R-parity violation can appear also in
supersymmetry breaking scalar potential:
\begin{equation} \label{VRp}
\Delta V = B_i L_i H_2 + m^2_{L_iH_1} L_i H_1^\dagger  +{\rm h.c.} \, ,
\end{equation}
where $H_1$ is the Higgs doublet with $Y=-1/2$.
In our basis, $B_i$ and $m^2_{L_iH_1}$
are related  by the condition of 
 vanishing sneutrino vacuum value,
$\langle \tilde{\nu}_i^* \rangle \propto B_i \langle H_2\rangle
+m^2_{L_i H_1}   \langle H_1\rangle =0$,
so the entire bilinear R-parity violation 
are described by  $\epsilon_i$ and $B_i$.

The basis described above is  not  the exact mass eigenbasis
of  fermions yet. If $\epsilon_i\neq 0$,
there can be a mass-mixing of order $\epsilon_i\mu$
between R-parity even and R-parity odd fermions \cite{nowa}.
Before taking into account of this mixing, we have four types
of fermion--scalar interaction vertices:
\beq
\label{vertices}
g\phi^*\psi\tilde{G}, 
\quad hH\psi\psi, \quad h\phi\psi\tilde{H},
\quad \lambda\phi\psi\psi,
\eeq
where the first three terms stand for  R-parity conserving 
gauge, Yukawa and Higgsino vertices,
respectively, and the last one does for R-parity violating Yukawa vertices.
Here $\psi$ and $H$ collectively  denote R-parity even matter fermions 
and Higgs scalars, respectively, and $\phi$, $\tilde{G}$ and $\tilde{H}$
are R-parity odd matter scalars, gauginos and Higgsinos, respectively.
When written in terms of the mass eigenstate fermions,
these couplings give rise to the {\it mixing-induced} vertices of the
following types:
\beq
\label{mixingvertices}
\epsilon g\phi^*\psi\psi, \quad  \epsilon h\phi\psi\psi,
\quad \epsilon\lambda\phi\psi\tilde{G}, 
\quad \epsilon\lambda\phi\psi\tilde{H} \, ,
\eeq
where the first and  second terms originate from the
R-parity even gauge and Yukawa vertices
in (\ref{vertices}), respectively, while
the other terms are  from the R-parity odd Yukawa
vertex in (\ref{vertices}).
Considering the processes  connecting 
these mixing-induced vertices with those of (\ref{vertices}),
one can get bounds on the products of bilinear and trilinear
couplings, which we will do explicitly
in the subsequent discussion.
There are also vertices induced by scalar mixing with $B_i\neq 0$ which
we do not consider here.  At this point, it is worthwhile to mention that 
the terms in (\ref{mixingvertices}) appear with 
$\langle \tilde{\nu}_i \rangle/\langle H_1 \rangle$ instead of $\epsilon_i$,
and with accordingly defined couplings $h$ and $\lambda$
in the different basis where the bilinear term $
\epsilon_i\mu L_iH_2$ in the superpotential is 
rotated away.
The most interesting feature of the mixing-induced 
vertices is that they have different chiral structure
compared to their  counterparts in (\ref{vertices})  
{\it unless} they do not include
the R-parity conserving Yukawa coupling $h$,
{\it i.e.} $\epsilon g\phi^*\psi\psi$
compared to  $\lambda\phi\psi\psi$ and also 
$\epsilon\lambda\phi\psi\tilde{G}$ compared to
$g\phi^*\psi\tilde{G}$.
This allows for instance that, when combined with
$g\phi^*\psi\tilde{G}$,
the mixing-induced $\epsilon\lambda\phi\psi\tilde{G}$ 
generates a dipole moment of light fermion
at one loop order {\it without} any insertion of 
the small Yukawa coupling of light fermion \cite{CCH}.

Let us write down explicitly the  mixing-induced vertices which are relevant
for our analysis. 
Ignoring the pieces involving the CKM matrix elements,
the R-parity odd part is given by
\begin{eqnarray} \label{indodd}
&&
g \Theta^{R*}_{iW} e^c_i 
[d_k \tilde{u}_k^* + e_k \tilde{\nu}_k^*]
+g \Theta^{L*}_{iW} e_i 
[u_k \tilde{d}_k^* + \nu_k \tilde{e}_k^*]  
+{\sqrt{2}g} [\Theta^{N*}_{iB}t_WY_f + \Theta^{N*}_{iW} T^3_f] \nu_i f\tilde{f}^*
\nonumber\\
&&-h_i^e \Theta^{L*}_{jH} e_j 
  [e_i^c \tilde{\nu}_i + \nu_i \tilde{e}_i^c]
-h_i^d \Theta^{L*}_{jH} e_j 
  [d_i^c \tilde{u}_i + u_i \tilde{d}_i^c]
+h_i^u \Theta^{R*}_{jH} e_j^c 
  [u_i^c \tilde{d}_i + d_i \tilde{u}_i^c] \nonumber \\
&&+h_i^e \Theta^{N*}_{jH_1} \nu_j 
  [e_i^c \tilde{e}_i + e_i \tilde{e}_i^c]
+h_i^d \Theta^{N*}_{jH_1} \nu_j 
  [d_i^c \tilde{d}_i + d_i \tilde{d}_i^c]
-h_i^u \Theta^{N*}_{jH_2} \nu_j 
  [u_i^c \tilde{u}_i + u_i \tilde{u}_i^c]
\end{eqnarray}
where $l_i=(\nu_i,e_i)$ and $q_i=(u_i,d_i)$ are the lepton and
quark doublets with the diagonal Yukawa couplings
$h_i^e$, $h_i^u$, $h_i^d$ with the Higgs doublets,
$e_i^c$, $u_i^c$, $d_i^c$ are the anti-lepton
and anti-quark singlets, and  the tilded fields stand for
their scalar superpartners.
Here $Y_f$ and $T^3_f$
denote the weak hypercharge and isospin of the fermion $f$,
$\Theta^{R}_{iW,H}$ ($\Theta^{L}_{iW,H}$) are the mixing elements between 
the charged lepton $e^c_i$ ($e_i$)
and the chargino $\tilde{W}^+,\tilde{H}^+$ ($\tilde{W}^-,\tilde{H}^-$),
 and  $\Theta^N_{iB,W,H_1,H_2}$ is the mixing element between the neutrino 
$\nu_i$ and the neutralino $\tilde{B}$,$\tilde{W}_3$,$\tilde{H}_1$,
$\tilde{H}_2$. 
The R-parity even part of the mixing-induced vertices
which are relevant for our analysis is given by
\beq \label{indeven}
\Theta^N_{i\chi}\lambda_{ijk} \chi^0 
[e_j\tilde{e}^c_k+ e^c_k \tilde{e}_j] 
+\Theta^L_{jW}\lambda_{ijk} \tilde{W}^- 
[\nu_i\tilde{e}^c_k+ e^c_k \tilde{\nu}_i] 
+\Theta^R_{kW}\lambda_{ijk} \tilde{W}^+ 
[\nu_i\tilde{e}_j+ e_j \tilde{\nu}_i]  
\eeq
where $\chi$ denotes the neutral gauginos $\tilde{B}$ and $\tilde{W}_3$.
There are also similar vertices involving $\lambda'_{ijk}$
which are not relevant for the discussion in this paper.
The detailed form of the mixing elements $\Theta$'s in Eqs.(\ref{indodd})
and (\ref{indeven}) can be found in Refs.~\cite{CCH,Lee}, 
however the following approximate expressions 
are enough for our purpose:
\begin{eqnarray} \label{THs}
&&\Theta^{R}_{iW} \approx \sqrt{2} {m^e_{i} \over M_{1/2}} 
               {M_W \over M_{1/2}} \epsilon_i c_\beta\,,\quad
\Theta^{L}_{iW} \approx  \sqrt{2} {M_W \over M_{1/2}} \epsilon_i^* c_\beta
      \nonumber\\
&&\Theta^{N}_{iB} \approx s_W {M_Z \over M_{1/2}} \epsilon_i^* c_\beta\,,
\quad \qquad
\Theta^{N}_{iW} \approx  -c_W {M_Z \over M_{1/2}} \epsilon_i^* c_\beta
       \nonumber\\ 
&&\Theta^{L}_{iH} \approx 2 s_\beta \left( {M_W\over M_{1/2}} \right)^2
  \epsilon_i^* c_\beta + \epsilon_i^* \,, \quad
\Theta^{R}_{iH} \approx {m_i^e \over c_\beta M_{1/2}} \epsilon_i c_\beta
      \nonumber \\
&&\Theta^{N}_{iH_1} \approx s_\beta {M_Z^2 \over \mu M_{1/2}} 
     \epsilon_i^* c_\beta \,, \quad \qquad
\Theta^{N}_{iH_2} \approx -c_\beta {M_Z^2 \over \mu M_{1/2}}
     \epsilon_i^* c_\beta 
\end{eqnarray}
where $m^e_i$ and $M_{1/2}$ denotes  the charged lepton mass and
the gaugino mass, respectively,
$M_W$ and $M_Z$ are the weak gauge boson masses,
$s_W=\sin\theta_W$ and $c_W=\cos\theta_W$ for the weak mixing
angle $\theta_W$,
$c_{\beta}=\cos\beta$ and $s_{\beta}=\sin\beta$ for
$\tan\beta=\langle H_2\rangle/\langle H_1\rangle$.

Combined with trilinear R-parity violation, 
bilinear R-parity violation induces  a left-right mixing 
of scalars which can contribute to the one-loop dipole moments 
of light fermions\cite{KK}, and thus
to $\mu\rightarrow e\gamma$.
{}From the superpotential (\ref{WZ}), one easily finds 
such type of mass mixing for squarks and sleptons:
\begin{equation} \label{LR}
(\epsilon^*_i \lambda_{ijk}\mu v_2)  \tilde{e}_j \tilde{e}^c_k
+ (\epsilon^*_i \lambda'_{ijk}\mu v_2)  \tilde{d}_j \tilde{d}^c_k
+h.c. \, ,
\end{equation}
where $v_2=\langle H_2\rangle$.
This left-right mixing  gives  rise to one-loop diagrams for 
$\mu\rightarrow e\gamma$ together with  two insertions of
the gauge vertices $g\phi^*\psi\tilde{G}$ without having any
insertion of light fermion mass.
We remark again that, 
in the basis where $L_iH_2$ in the superpotential is rotated away,
the similar sfermion mixing terms arise from trilinear soft-terms 
with the coefficients $\langle \tilde{\nu}_i \rangle \lambda_{ijk}$ or
$\langle \tilde{\nu}_i \rangle \lambda'_{ijk}$.

We are now ready to consider the processes induced by 
diagrams connecting the mixing-induced 
vertices in (\ref{mixingvertices})
with the conventional vertices in (\ref{vertices}).
The processes  $K_L\rightarrow e_i\bar{e}_j$,
$K^+\rightarrow\pi^+\nu_i\nu_j$ and also the $\mu$--$e$
conversion in nuclei can be triggered by tree diagrams involving
both bilinear and trilinear R-parity violations.
(See Fig. 1.)
It is in fact straightforward to obtain the bounds on 
$\epsilon\lambda^{\prime}$ from these processes.
The diagrams of Fig. 1 correspond to the diagrams 
which have been considered in Refs.\cite{CR,AG,huitu,Faessler,CH} 
to obtain the bounds on $\lambda^{\prime}\lambda^{\prime}$, but
here one $\lambda'$ is replaced by appropriate mixing-induced  coupling  
in (\ref{indodd}) while properly taking into account the necessary 
chirality flip.
The  resulting bounds  are summarized in Table I. 
Here the bounds on $\lambda^{\prime}_{ijk}\epsilon_lc_{\beta}$ are 
from diagrams
with a mixing-induced vertex of the form $\epsilon g\phi^*\psi\psi$,
while the bounds  on $\lambda^{\prime}_{ijk}\epsilon_l/c_{\beta}$ are from
those with  $\epsilon h\phi\psi\psi$.

Let us now consider $\mu\rightarrow e\gamma$. 
As one can see in Fig. 2, the  mixing-induced  vertices of 
Eq.(\ref{indeven})
can generate one-loop  dipole moments of light fermions
without having the insertion of  light fermion mass.
The required chirality flip
is provided by the gaugino mass.
Indeed, similar diagrams have been used to generate the leading 
contribution to the  electric dipole moments of the electron and
neutron in Ref.~\cite{CCH}.  
On the other hand, there are other type of one-loop diagrams for
dipole moments without any light fermion mass.
This type of diagrams  involve the left-right sfermion mixing of 
Eq. (\ref{LR}) together with two insertions of the
standard gaugino vertices.
(See the diagram of Fig. 2b.)
Note that there are two chirality flips in the internal lines
of this type of diagram,
one is the left-right
sfermion mixing and the other is the gaugino mass mixing.
For $\mu \to e \gamma$, there is only one
diagram with $\tilde{B}$--$\tilde{W}_3$  mixing, giving somewhat
smaller contributions.

The decay width of $\mu\rightarrow e\gamma$ 
is given by
\begin{equation}
 \Gamma(\mu \to e\gamma) = {m_\mu^3 \over 16\pi} 
 \left( |{\cal B}_{L}|^2 + |{\cal B}_{R}|^2\right) \, ,
\end{equation}
for the transition dipole moments
\beq
\frac{i}{2}\bar{e}\, \sigma^{\mu\nu} 
\left({\cal B}_R P_R+{\cal B}_L P_L\right) \mu F_{\mu\nu} \, ,
\eeq
where $P_{L,R}=\frac{1}{2}(1\pm \gamma_5)$.
The dominant contributions to  ${\cal B}_{R,L}$ from the diagrams
in Fig. 2  are found to be
\begin{eqnarray}
{\cal B}_{R}  = {eg \over 16\pi^2} \left\{ \right. && \left.
 { \Theta^L_{jW} \lambda_{1 j 2} \over m_{\tilde{\nu}_1}}
 G_f(m_{\tilde{W}}; m_{\tilde{\nu}_1}) \right.
       \nonumber\\
&&- {1\over\sqrt{2}} 
   { \lambda_{1j2} \over m_{\tilde{e}_1}}
\left[ \Theta^N_{jW}
 G_s(m_{\tilde{W}};m_{\tilde{e}_1}) 
+ t_W \Theta^N_{jB} 
 G_s(m_{\tilde{B}};m_{\tilde{e}_1}) \right] \\
&& +\sqrt{2} t_W  \left.
   { \Theta^N_{jB} \lambda_{1 j 2} \over m_{\tilde{e}^c_2}}
 G_s(m_{\tilde{B}}; m_{\tilde{e}^c_2}) \right\} \nonumber \\
+{eg^2 t_W \over 16\pi^2}&&
N_{B\chi_n}N_{W\chi_n} {\epsilon^*_j \lambda_{jik}
     \mu v_2 \over m_{\tilde{e}_i}m_{\tilde{e}^c_k}}
 H_s(m_\chi^0; m_{\tilde{e}_i},m_{\tilde{e}^c_k}) \, ,\nonumber \\
{\cal B}_{L} = (1 \leftrightarrow && 2)\quad
         {\rm exchange} \quad {\rm of} \quad
        {\cal B}_{R} \, .
\end{eqnarray}
Here $N_{B\chi_n}$ and $N_{W\chi_n}$ are the 
diagonalization matrix bringing the neutralinos
$(\tilde{B},\tilde{W}_3,\tilde{H}^0_1,\tilde{H}^0_2)$
into the mass eigenstates $\chi_n$,
and the loop functions $G_f$,  $G_s$
and $H_s$  are given by
\bea
G_f(m_f; m_\alpha)&=&
t\left[{t^2-3 \over 2(1-t^2)^2} - {2\ln t \over (1-t^2)^3}\right] \,,
\nonumber \\
G_s(m_f;m_\alpha)&=&  
t\left[{t^2+1 \over 2(1-t^2)^2} + {2 t^2 \ln t \over (1-t^2)^3}\right] \,,
\nonumber \\
H_s(m_f;m_\alpha, m_\beta) &=&
{m_\alpha m_\beta \over m_\alpha^2 - m_\beta^2}
\left[ {1\over m_\alpha} G_s(m_f; m_\alpha) 
- {1\over m_\beta} G_s(m_f; m_\beta) \right] \,,
\nonumber 
\eea
where $t=m_f/m_{\alpha}$.

Applying the experimental data \cite{pdg}
${\rm BR}(\mu \to e\gamma) < 4.9\times10^{-11}$ for the
above transition dipole moment,
one can derive bounds on the certain components
of $\lambda_{ijk}\epsilon^*_j$
for reasonable range of superparticle spectrums.
The results are summarized in Table II.
In the absence of slepton generation mixing, $\mu\rightarrow
e\gamma$ would provide bounds only on the components
$\lambda_{1j2\epsilon^*_j}$ and $\lambda_{2j1}\epsilon^*_j$.
However the basis that we are working on is not the mass eigenbasis
of sfermions, and thus there can be slepton generation mixing
induced by the off-diagonal elements $\Delta^{AB}_{ij}$
of the slepton mass-squared matrix. 
(Here $A,B=L,R$ denote the chirality of the corresponding sleptons.)
If we allow such slepton-mixing,
other components of $\lambda_{ijk}\epsilon^*_j$
can be constrained also by $\mu\rightarrow e\gamma$  
as in Table II.

Any lepton number violating interactions
are  highly constrained by neutrino masses.
It is thus quite tempting to compare the constraints
of Tables I and II with
the constraints from neutrino masses.
As mentioned before, we take the upper bound on 
each component of neutrino mass matrix  to be 1 eV \cite{haug},
although it depends
on the interpretation of Super-Kamiokande data and also
on the possible existence of light sterile neutrinos.
Assuming a simple flavor structure
of $\lambda_{ijk}$ and $\lambda^{\prime}_{ijk}$,
strong bounds on $\lambda\lambda$, $\lambda^{\prime}\lambda^{\prime}$,
and $\epsilon\epsilon$ were derived from neutrino oscillation
data \cite{abada}. Here we briefly discuss this issue
without any assumption on the flavor structure of $\lambda_{ijk}$ 
and $\lambda^{\prime}_{ijk}$, and compare the results with
the constraints of  Tables I and II.
We will see that the constraints of Tables I and II
from flavor changing processes 
are complementary to the constraints from neutrino masses.

In the basis with vanishing sneutrino vacuum values,
R-parity violation are described by  
$\{\epsilon_i, B_i, \lambda_{ijk},
\lambda^{\prime}_{ijk}\}$. 
Here we are interested in the neutrino masses
associated with $\epsilon_i$, $\lambda_{ijk}$ and
$\lambda^{\prime}_{ijk}$, and thus ignore the contribution
involving $B_i$. 
Then tree-diagrams with two insertions of the lepton-neutralino mixing
give rise to
\beq
\label{treemass}
(m_{\nu})^{\rm tree}_{ij}\sim
\frac{M_Z^2 c^2_\beta}{M_{1/2}}\epsilon_i\epsilon_j,
\eeq
while one-loop diagrams with two insertions of $\lambda_{ijk}$
or of $\lambda^{\prime}_{ijk}$ 
give 
\bea
\label{loopmass}
(\delta_1 m_{\nu})^{\rm loop}_{ij}&\sim&
\frac{3}{8\pi^2}\frac{m^{d}_km^{d}_l}{m^2_{\tilde{q}}}
\lambda^{\prime}_{ikl}\lambda^{\prime}_{jlk}
[A^{d*}+\mu t_\beta ],
\nonumber \\
(\delta_2 m_{\nu})^{\rm loop}_{ij}&\sim&
\frac{1}{8\pi^2}\frac{m^e_km^e_l}{m^2_{\tilde{l}}}
\lambda_{ikl}\lambda_{jlk}[A^{e*}
+\mu t_\beta],
\eea
where $m^{d}_i$ and $m^e_i$ are the down-type quark and charged
lepton masses,  $A^{d}$ and $A^e$ are the soft $A$-parameter
for the trilinear couplings $H_1QD^c$ and $H_1LE^c$, $m_{\tilde{q}}$
and $m_{\tilde{l}}$ are the squark and slepton masses, and
$c_{\beta}=\cos\beta$ and $t_{\beta}=\tan\beta$.
There are also one-loop diagrams giving neutrino masses
proportional to the products of bilinear and trilinear couplings\footnote{
Such one-loop mass has been noticed  also in Ref.~\cite{cch1} 
within the different parameter basis with
$\epsilon_i=0$ but $\langle\tilde{\nu}_i\rangle\neq 0$.}.
Transparent description of such one-loop diagrams can be made by the use
of mass insertion approximation \cite{davidson}.  
The mass insertion by $\epsilon_i$ can appear  either on the external
fermion (neutralino/neutrino) line or on the internal chargino/charged lepton
line.  For each of these two types, 
there are diagrams in which  one vertex is given by
R-parity odd Yukawa coupling ($\lambda$ or $\lambda'$) and the other  by
R-parity even gauge or Yukawa ($h^e$ or $h^d$) coupling. It turns out that
the diagrams with $h^e$ or $h^d$ are suppressed by two more powers
of $m^e_i$ or $m^d_i$ which come from the chirality flips in 
the internal lepton/slepton or quark/squark lines.
These one-loop diagrams give the following contributions \cite{davidson}:
\bea
\label{anotherloopmass}
(\delta_3 m_{\nu})^{\rm loop}_{ij} &\sim& 
\frac{3g}{16\pi^2} \frac{M_Z}{\sqrt{2} M_{1/2}} c_\beta
(\epsilon_i\lambda^{\prime}_{jkk}+\epsilon_j\lambda^{\prime}_{ikk})
m^d_k \nonumber\\
&&+ \frac{3}{16\pi^2} \frac{M_Z^2}{\mu M_{1/2}} s_\beta c_\beta
( \epsilon_i\lambda^{\prime}_{jkk}+\epsilon_j\lambda^{\prime}_{ikk})
{h^d_k (m^d_k)^2 \over m^2_{\tilde{q}}} [A^{d*}+\mu t_\beta]
\nonumber \\
(\delta_4 m_{\nu})^{\rm loop}_{ij} &\sim& 
\frac{g}{16\pi^2} \frac{M_Z}{\sqrt{2} M_{1/2}} c_\beta
( \epsilon_i\lambda_{jkk}+\epsilon_j\lambda_{ikk}) m^e_k 
\nonumber\\
&&+ \frac{1}{16\pi^2} \frac{M_Z^2}{\mu M_{1/2}} s_\beta c_\beta
( \epsilon_i\lambda_{jkk}+\epsilon_j\lambda_{ikk})
{h^e_k (m^e_k)^2\over m^2_{\tilde{l}}} [A^{e*}+\mu t_\beta] 
\nonumber\\
&&+ \frac{1}{16\pi^2} 
(\epsilon_k \lambda_{ijk})
{( h^e_i m^e_i- h^e_j m^e_j) m^e_k \over m^2_{\tilde{l}}}
[A^{e*}+\mu t_\beta] 
\eea
where $h^d_i$ and $h^e_i$ are the diagonalized quark and lepton 
Yukawa couplings generating the masses $m^d_i$ and $m^e_i$.
Requiring that all of these contributions
are smaller than $1$ eV, we obtained upper bounds of
several combinations of couplings, which are summarized
in Table III.
The $t_\beta$ dependence in Table III
 comes from the sfermion mixing mass term 
involving $\mu t_\beta$.

Note that the tree level neutrino mass (\ref{treemass}) is not
suppressed by small fermion mass, and thus leads to a strong bound
of order $10^{-5}\sim 10^{-6}$ on $\epsilon_i$.
Since all the loop-induced neutrino masses are suppressed by 
some powers of small fermion masses as well as the loop suppression factor,
the light generation components of R-parity violating couplings
are not severely constrained by neutrino masses.

We have found that the consideration of flavor changing
decays of the neutral kaon and also of $\mu$--$e$ conversion
in nuclei put bounds of order $10^{-4}\sim 10^{-5}$ on  
$\lambda^{\prime}_{ijk}\epsilon_l c_\beta^{\pm 1}$
as summarized in Table I.
Also $\mu\rightarrow e\gamma$ constrain $\lambda_{i12}\epsilon^*_i$ and
$\lambda_{i21}\epsilon^*_i$ to be less than about $10^{-7}$
as in Table II.
Obviously, there are  large range of parameters satisfying
the bounds of Table III, but excluded by   the bounds of Table II.  
Thus $\mu\rightarrow e\gamma$ 
provides  meaningful additional constraints on R-parity violation
over the constraints from neutrino masses.
It is also true that large range of parameters satisfying
the bounds of Table II can be excluded by the bounds of Table III.
All bounds of Table I would be trivially satisfied for 
small $\tan\beta$ if the bounds of Table III are imposed.
We still note that if $\tan\beta$ is greater than 10, some bounds
of Table I provide meaningful additional constraints on R-parity
violation over the bounds from neutrino masses.
Furthermore, if a light sterile neutrino is introduced,
it may not be necessary that all of the three weak doublet 
neutrinos have masses smaller than $1$ eV to explain the 
neutrino oscillation data, which would allow some of the bounds
of Table III relaxed.

To conclude, we have derived strong bounds on certain products
of the bilinear and trilinear R-parity violating couplings from 
flavor changing decays of the neutral kaon, $\mu$--$e$
conversion in nuclei, and also $\mu \to e\gamma$.
We also discussed the constraints on R-parity violation
from neutrino masses under the assumption that existing 
neutrino data imply
that all the neutrino masses are smaller than 1 eV.
Wide range of parameter space satisfying the
constraints from neutrino masses can be excluded
by the bounds from flavor changing processes, and 
vice versa, thus these two sets of 
constraints are complementary to each other.

\bigskip

{\bf Acknowledgement}

This work is supported by Seoam Foundation (K.C.) and by the KOSEF grant
No.1999-2-111-002-5 and BK21 project of the Ministry of Education
(K. C. and K. H.)



\renewcommand{\arraystretch}{1}
\begin{table}
\begin{tabular}{|c|c|c|l|}
\quad  process 
\quad & coupling 
\quad &  exchanged scalar mass
\quad & upper bound   \\ \hline
$K_L \to \mu\bar{\mu} \cite{CR}:$ &
  $\lambda'_{221} \epsilon_2/c_\beta$ & 
  $m^2_{\tilde{u}_{2}} $ &$ 5.7\times10^{-4} x_s^2$ \\ 
& $\lambda'_{212} \epsilon_2/c_\beta$ &
  $m^2_{\tilde{\nu}_{2}}$ & $6.8\times10^{-4} x_s^2 $ \\ \hline
$ K_L \to e\bar{\mu},\, \mu\bar{e} \cite{CR}:$ &
  $ \lambda'_{121} \epsilon_2/c_\beta,
    \lambda'_{221} \epsilon_1/c_\beta $ &
  $ m^2_{\tilde{u}_{2}}$ & $3.4\times10^{-5} x_s^2$ \\
& $ \lambda'_{112} \epsilon_2 c_\beta $ &
  $m^2_{\tilde{u}_{1}}, m^2_{\tilde{\nu}_{1}}$ &
  $ 3.1\times10^{-5} x_s^2 x_g^2$ \\
& $ \lambda'_{221,212} \epsilon_1/c_\beta $ &
  $ m^2_{\tilde{\nu}_{2}} $ & $ 4.1\times10^{-5} x_s^2 $ \\ \hline
$ \mu + Ti \to e + Ti \cite{Faessler}:$ &
  $ \lambda'_{111} \epsilon_2c_\beta $ &
  $ m^2_{\tilde{\nu}_{1}}, m^2_{\tilde{u}_{1}}$ & 
  $ 1.1\times10^{-5}  x_s^2 x_g^2$ \\
& $ \lambda'_{122} \epsilon_2/c_\beta $ &
  $ m^2_{\tilde{u}_{2}} $ & $4.9\times10^{-4}  x_s^2$ \\
& $ \lambda'_{211} \epsilon_1/c_\beta $ &
  $ m^2_{\tilde{\nu}_{2}} $ & $ 7.4\times10^{-6} x_s^2 $ \\
& $ \lambda'_{222} \epsilon_1/c_\beta $ &
  $ m^2_{\tilde{\nu}_{2}} $ & $1.4\times10^{-5} x_s^2 $ \\ \hline
$ K^+ \to \pi^+\nu\bar{\nu} \cite{AG,CR}: $ &
  $ \lambda'_{i12,i21} \epsilon_l c_\beta $ &
  $ m^2_{\tilde{d}_{1,2}} $ & $1.2\times10^{-4} x_s^2 x_g$ \\
\end{tabular}

\bigskip

\caption {
Bounds on the products of trilinear and bilinear couplings
from various flavor changing processes induced by tree
diagrams.
Here $x_s \equiv {m_{\tilde{f}}\over 100 {\rm GeV}}$,
 $x_g \equiv {M_{1/2}\over 100 {\rm GeV}}$
 and $c_\beta = \cos\beta$.
}
\vspace{1cm}
\end{table}

\renewcommand{\arraystretch}{1}
\begin{table}
\begin{tabular}{|c|c|}
\quad\quad\quad\quad\quad\quad\quad\quad
coupling 
\quad\quad\quad\quad\quad\quad\quad\quad
& upper bound 
\quad\quad\quad\quad\quad\quad\quad\quad
\\ \hline
$ \lambda_{1j2} \epsilon_j^* c_\beta$ &
 $ 1.5\times10^{-7} x_sx_g $ \\
$ \lambda_{2j1} \epsilon_j^* c_\beta$ &
 $ 1.5\times10^{-7} x_sx_g $ \\
$ \delta^{LL}_{1i} \lambda_{i j2} \epsilon_j^* c_\beta$ &
 $ 1.5\times10^{-7} x_sx_g $ \\
$ \delta^{LL}_{2i} \lambda_{ij1} \epsilon_j^* c_\beta$ &
 $ 1.5\times10^{-7} x_sx_g $ \\
$ \delta^{RR}_{1k} \lambda_{2jk} \epsilon_j^* c_\beta$ &
 $ 1.0\times10^{-6} x_sx_g $ \\
$ \delta^{RR}_{2k} \lambda_{1jk} \epsilon_j^* c_\beta$ &
 $ 1.0\times10^{-6} x_sx_g $ \\
$ \lambda_{1j2} \epsilon_j^* {\mu v_2 \over \tilde{m}^2} $ &
 $ 5.5\times10^{-6} x_s^2 x_g^2 $ \\
$ \lambda_{2j1} \epsilon_j^* {\mu v_2 \over \tilde{m}^2} $ &
 $ 5.5\times10^{-6} x_s^2 x_g^2 $
\end{tabular}

\bigskip

\caption{
Bounds on the products of trilinear and bilinear couplings from
$\mu \to e \gamma$. 
Here $x_s \equiv {m_{\tilde{f}}\over 100 {\rm GeV}}$
, $x_g \equiv {M_{1/2}\over 100 {\rm GeV}}$, $c_\beta = \cos\beta$,
and $\delta^{AB}_{ij}=\Delta^{AB}_{ij}/m_{\tilde{l}_i}
m_{\tilde{l}_j}$ for the off-diagonal elements
$\Delta^{AB}_{ij}$ of the slepton mass-squared matrix 
with $A,B=L,R$ denoting the
chirality of sleptons.
In all these bounds, 
$c_\beta$ can be replaced  by ${M_W\over M_{1/2}}$.
}
\vspace{1cm}
\end{table}

\renewcommand{\arraystretch}{1}
\begin{table}
\begin{tabular}{|c|c|}
\quad\quad\quad\quad\quad\quad\quad\quad
coupling 
\quad\quad\quad\quad\quad\quad
& upper bound 
\quad\quad\quad\quad\quad\quad\quad\quad
\\ \hline
$ \epsilon_i$ & 
$ 3\times10^{-6} x_s^{1/2}/c_\beta $  \\
$ \lambda_{ikl}\lambda_{jlk}y^e_ky^e_l $ &
$ 2\times10^{-6} x_s/t_\beta$ \\
$ \lambda^{\prime}_{ikl}\lambda^{\prime}_{jlk}y^{d}_ky^{d}_l $ & 
$ 10^{-7} x_s/t_\beta $ \\
$ \epsilon_i\lambda_{jkk}y^e_k $ &
$ 10^{-7} x_s/c_\beta $ \\
$ \epsilon_i\lambda^\prime_{jkk}y^d_k $ & 
$ 2\times10^{-8} x_s/c_\beta $ \\
$ \epsilon_k\lambda_{ijk}y^e_k[(y^e_i)^2-(y^{e}_j)^2] $ &
$ 5\times10^{-4} x_s /t_\beta $ \\
$ \epsilon_i\lambda_{jkk}(y^e_k)^3 $ &
$ 5\times10^{-4}x_s^3 $ \\
$ \epsilon_i\lambda^{\prime}_{jkk}(y^d_k)^3 $ &
$ 10^{-5} x_s^3 $ 
\end{tabular}

\bigskip

\caption{
Constraints on R-parity violation from the condition that
all elements of the $3\times 3$ neutrino mass matrix
are smaller than 1 eV,  which would be necessary to account
for the neutrino oscillation data.
Here $y^e_i\equiv m^e_i/m_\tau$, $y^d_i\equiv m^d_i/m_b$ 
and $c_\beta = \cos\beta, t_\beta = \tan\beta$
, while $x_s=M_{SUSY}/100{\rm GeV}$ for the
SUSY mass scales 
$M_{SUSY}=\{M_{1/2},m_{\tilde{f}}, \mu, A^{d,e}\}$. 
}
\vspace{1cm}
\end{table}


\begin{figure}
\begin{center}
\begin{picture}(460,460)(0,0)

\SetOffset(0,280)

\Text(100,-20)[]{(a)}

\ArrowLine(20,130)(100,130)
\ArrowLine(180,130)(100,130)
\ArrowLine(100,30)(20,30)
\ArrowLine(100,30)(180,30)
\DashArrowLine(100,30)(100,130){5}

\Text(30,18)[]{$d_j^c$}
\Text(30,142)[]{$d_k^c$}
\Text(170,18)[]{$\ell_l$}
\Text(170,142)[]{$\ell_i$}
\Text(110,82)[]{$\tilde{u}_j$}

\Text(100,142)[]{$\lambda'_{ijk}$}
\Text(100,18)[]{$\epsilon_l^* h_j^d$}

\SetOffset(0,50)

\Text(100,-20)[]{(c)}

\ArrowLine(20,130)(100,130)
\ArrowLine(180,130)(100,130)
\ArrowLine(100,30)(20,30)
\ArrowLine(100,30)(180,30)
\DashArrowLine(100,30)(100,130){5}

\Text(30,18)[]{$u_k$}
\Text(30,142)[]{$u_j$}
\Text(170,18)[]{$\ell_l$}
\Text(170,142)[]{$\ell_i$}
\Text(110,82)[]{$\tilde{d}_k^c$}

\Text(100,142)[]{$\lambda'_{ijk}$}
\Text(100,18)[]{$\epsilon_l^* h_k^d$}

\SetOffset(230,280)

\Text(100,-20)[]{(b)}

\ArrowLine(20,30)(70,80)
\ArrowLine(20,130)(70,80)
\ArrowLine(130,80)(180,30)
\ArrowLine(130,80)(180,130)
\DashArrowLine(130,80)(70,80){5}

\Text(30,18)[]{$d_j$}
\Text(30,142)[]{$d_k^c$}
\Text(170,18)[]{$\ell_l$}
\Text(170,142)[]{$\ell_i$}
\Text(100,92)[]{$\tilde{\nu}_i$}

\Text(50,80)[]{$\lambda'_{ijk}$}
\Text(152,80)[]{$\epsilon_l^* h_k^d$}

\SetOffset(230,50)

\Text(100,-20)[]{(d)}

\ArrowLine(20,130)(100,130)
\ArrowLine(180,130)(100,130)
\ArrowLine(20,30)(100,30)
\ArrowLine(180,30)(100,30)
\DashArrowLine(100,30)(100,130){5}

\Text(30,18)[]{$d_j$}
\Text(30,142)[]{$d_k^c$}
\Text(170,18)[]{$\nu_l$}
\Text(170,142)[]{$\nu_i$}
\Text(110,82)[]{$\tilde{d}_j$}

\Text(100,142)[]{$\lambda'_{ijk}$}
\Text(100,18)[]{$\epsilon_l^* g,g'$}

\end{picture}

\end{center}
\caption{Important diagrams for $K_L\to\mu\bar{\mu}$,
$K_L\to e\bar{\mu},\, \mu\bar{e}$ (a,b),
$\mu + T_i \to e + T_i$ (a,b,c)
and $K^+ \to \pi^+\nu\bar{\nu}$ (d).}
\end{figure}

\begin{figure}
\begin{center}
\begin{picture}(300,560)(0,0)

\SetOffset(0,400)

\ArrowLine(10,50)(70,50)
\ArrowLine(150,50)(70,50)
\Line(145,45)(155,55) 
\Line(155,45)(145,55)
\ArrowLine(150,50)(230,50)
\ArrowLine(290,50)(230,50)

\DashArrowArcn(150,50)(80,180,0){5}

\Text(20,38)[]{$e$}
\Text(110,38)[]{$\tilde{W}^+$}
\Text(190,38)[]{$\tilde{W}^-$}
\Text(280,38)[]{$\mu^c$}

\Text(150,145)[]{$\tilde{\nu}_e$}

\Text(70,38)[]{$g$}
\Text(230,38)[]{$\epsilon_j^* c_\beta \lambda_{1j2}$}

\SetOffset(0,240)

\ArrowLine(10,50)(70,50)
\ArrowLine(150,50)(70,50)
\Line(145,45)(155,55) 
\Line(155,45)(145,55)
\ArrowLine(150,50)(230,50)
\ArrowLine(290,50)(230,50)

\DashArrowArc(150,50)(80,0,180){5}

\Text(20,38)[]{$e$}
\Text(110,38)[]{$\tilde{B}$}
\Text(190,38)[]{$\tilde{B}$}
\Text(280,38)[]{$\mu^c$}

\Text(150,145)[]{$\tilde{\mu}^c$}

\Text(70,38)[]{$-\epsilon_j^* c_\beta \lambda_{1j2}$}
\Text(230,38)[]{$g'$}

\Text(150,-10)[]{\rm (a)}

\SetOffset(0,0)

\ArrowLine(10,50)(70,50)
\ArrowLine(150,50)(70,50)
\Line(145,45)(155,55) 
\Line(155,45)(145,55)
\ArrowLine(150,50)(230,50)
\ArrowLine(290,50)(230,50)

\DashArrowArc(150,50)(80,0,90){5}
\DashArrowArcn(150,50)(80,180,90){5}
\Vertex(150,130){4}

\Text(20,38)[]{$e$}
\Text(110,38)[]{$\tilde{B}$}
\Text(190,38)[]{$\tilde{W}^3$}
\Text(280,38)[]{$\mu^c$}

\Text(75,115)[]{$\tilde{e}$}
\Text(225,115)[]{$\tilde{\mu}^c$}
\Text(150,145)[]{$\epsilon_j^* \lambda_{j12}\mu v_2$}

\Text(70,38)[]{$g$}
\Text(230,38)[]{$g'$}

\Text(150,10)[]{\rm (b)}

\end{picture}
\end{center}
\bigskip
\caption{Important diagrams for the dipole moment
${\cal B}_R$ of $\mu \to e\gamma$ decay.
Here, the cross denotes the gaugino mass insertion 
and the blop denotes the scalar mass mixing.
Diagrams for ${\cal B}_L$ 
can be obtained by changing $e$ and $\mu$ each other and
also $\lambda_{1j2}$ by $\lambda_{2j1}$.
}
\end{figure}


\begin{thebibliography}{99}
%
\def\plb#1#2#3{Phys.\ Lett.\       {\bf B#1}, #2 (#3)}
\def\npb#1#2#3{Nucl.\ Phys.\       {\bf B#1}, #2 (#3)}
\def\prd#1#2#3{Phys.\ Rev.\        {\bf D#1}, #2 (#3)}
\def\prl#1#2#3{Phys.\ Rev.\ Lett.\ {\bf #1},  #2 (#3)}
\def\mpl#1#2#3{Mod.\ Phys.\ Lett.\ {\bf A#1}, #2 (#3)}
\def\rep#1#2#3{Phys.\ Rep.\        {\bf #1},  #2 (#3)}
\def\sci#1#2#3{Science             {\bf #1},  #2 (#3)}
\def\astro#1#2#3{Astrophys.\ J.\   {\bf #1},  #2 (#3)}
%

\bibitem{review} For reviews, see for instance
H. Dreiner, hep-ph/9707435;
G. Bhattacharyya, hep-ph/9709395.
\bibitem{CR}
D.~Choudhury and P.~Roy, \plb{378}{153}{1996}.

\bibitem{AG}
K.~Agashe and M.~Graesser, \prd{54}{4445}{1996}.

\bibitem{huitu}
K. Huitu, J. Maalampi, M. Raidal and A. Santamaria, \plb{430}{355}{1998};
J.E. Kim, P. Ko and D.-G. Lee,  \prd{56}{100}{1997}.

\bibitem{Faessler}
A.~Faessler, T.S.~Kosmas, S.~Kovalenko and J.D.~Vergados, 
hep-ph/9904335.

\bibitem{CH}
B. de Carlos and P.L. White, \prd{54}{3427}{1996};
M. Chaichian and K. Huitu, \plb{384}{157}{1996}.

\bibitem{CCH}
K.~Choi, E.J.~Chun and K.~Hwang, hep-ph/0004101.

\bibitem{KK}
Y.-Y.~Keum, O.C.W.~Kong, hep-ph/0004110.

\bibitem{hs}
L.  Hall and M. Suzuki, \npb{231}{419}{1984}.

\bibitem{haug}
O. Haug {\it et al.}, \npb{565}{38}{2000}. 

\bibitem{abada}
A. Abada and M. Losada, hep-ph/9908352.

\bibitem{Lee}
E.J. Chun and J.S. Lee, \prd{60}{075006}{1999}.

\bibitem{skam}  
The Super-Kamiokande Collaboration, Y. Fukuda, {\it et al.},
\prl{81}{1562}{1998}.

\bibitem{yuval} 
M. Bisset, O.C.W. Kong, C. Macesanu and L.H. Orr, \plb{430}{274}{1998};
Y. Grossman and H.E. Haber, \prd{59}{093008}{1999}.

\bibitem{banks}
T. Banks, Y. Grossman, E. Nardi and Y. Nir, \prd{52}{5319}{1995};
H. -P. Nilles and N. Polonsky, \npb{484}{33}{1997};
S. Davidson and J. Ellis, \plb{390}{210}{1997};
\prd{56}{4182}{1997};
S. Davidson, \plb{439}{63}{1998};
J. Ferrandis, \prd{60}{095012}{1999};
S. Davidson, M. Losada and N. Rius, hep-ph/9911317;
Y. Grossman and H. E. Haber, preprint in preparation.

\bibitem{nowa}
A.S. Joshipura and M. Nowakowski, \prd{51}{2421}{1995};
M. Nowakowski and A. Pilaftsis, \npb{461}{19}{1996}.

\bibitem{pdg}
Particle Data Group, Euro.\ Phys.\ J.\ {\bf C1}, 1 (1998).

\bibitem{cch1}
K. Choi, E.J. Chun and K. Hwang, \prd{60}{031301}{1999};
D. E. Kaplan and A. E. Nelson, JHEP 0001, 033 (2000).

\bibitem{davidson}
S. Davidson and M. Losada, hep-ph/0005080;
See for the full descprition, 
E.J. Chun and S.K. Kang, \prd{61}{075012}{2000}. 

\end{thebibliography}
\end{document}